# EmpowerAbility: A portal for employment & scholarships for differently-abled

HIMANSHU RAJ
School of Computing
SRM Institute of Science and Technology
Kattankulathur, Tamil Nadu, India
hr0065@srmist.edu.in

SHUBHAM KUMAR
School of Computing.
SRM Institute of Science and Technology
Kattankulathur, Tamil Nadu, India
sp9787@srmist.edu.in

DR. J. KALAIVANI
School of Computing.
SRM Institute of Science and Technology
Kattankulathur, Tamil Nadu, India
kalaivaj@srmist.edu.in

*Abstract—*

**The internet has become a vital resource for job seekers in today's technologically advanced world, particularly for those with impairments. They mainly rely on internet resources to find jobs that fit their particular requirements and skill set. Though some disabled candidates receive prompt responses and job offers, others find it difficult to traverse the intricate world of job portals, the efficacy of this process frequently varies. This discrepancy results from a typical error: a failure to completely comprehend and utilize the accessibility features and functions that can significantly expedite and simplify the job search process for people with impairments.**

**This project is a job and scholarship portal that empowers individuals with diverse abilities. Through inspiring success stories, user-centric features, and practical opportunities, it fosters resilience and inclusivity while reshaping narratives. This platform's dual-pronged strategy instills pride and offers real-world solutions, making a lasting impact on the lives it touches.**

## I. INTRODUCTION

In India, a country with a population of more than 1.3 billion, there is a sizeable demographic of about 2.1% of persons with special needs, or roughly 2 crore people. These extraordinary people are brimming with undiscovered brilliance, innumerable original ideas, and wealth of unrealized potential. Nevertheless, despite the size of this population, a crucial question still remains: why do possibilities for our friends with disabilities frequently stay elusive? This query served as the catalyst for the development of this project.

The conspicuous lack of interaction and accessibility to online information for people with special needs is one of the most important challenges we discovered during our investigation. Even in the information-rich digital age, this demographic still suffers significant access challenges. It is noteworthy that government websites, which frequently include important information, are cluttered with complications that make it difficult to traverse them, let alone code them to be accessible.

The project is incorporated with cutting-edge features to make it even more approachable and user-friendly. Think about India's 8.8 million visually handicapped citizens. Voice-controlled navigation and an auto-read capability has been added to meet their specific demands, ensuring that the platform is genuinely inclusive.

In this research paper, we delve into the essence of this project, investigating its ground-breaking approaches and effects, and illuminating how it helps our friends with special needs bridge the gap between potential and opportunity by turning statistics into success stories.

## II. MOTIVATION

Unmet Need for inclusion: The project was inspired by the conspicuous absence of accessibility and inclusion for people with disabilities, especially in the context of online possibilities and services.[2] There is a moral and cultural obligation to close this gap and guarantee equitable access to opportunities given the sizeable population of individuals with special needs in India and their enormous untapped potential.

High Unemployment Rates for People with Special Needs: Studies show that people with special needs have shockingly high unemployment rates in India, and statistics show that a sizeable section of this population is still unemployed despite their talents. [2] People's employment is insecure; this is evidenced by things like regular job changes, a reliance on blue-collar jobs, inadequate qualifications, and a lack of computer literacy. Another significant worry is the frequency of job insecurity in the industry, which is influenced by participants' perceptions of their limited employability and career anxiety. [1] This provides a strong incentive to develop a platform that not only points out opportunities, but also removes the structural obstacles to work and education.

Potential for Transformative Impact: The project is motivated by the conviction that maximizing the potential of people with special needs can have significant societal repercussions. These individuals face various stereotypical discriminations on the basis of their disability, in the society. [5] The project seeks to empower individuals while also challenging society preconceptions and stereotypes, ultimately influencing narratives around disability. It does this through offering chances and presenting success stories.

Technological Innovation: The project's motive is given a technological twist by the use of cutting-edge tools like web scraping and voice-enabled text-to-speech functionalities driven by AI. These technologies are used to improve accessibility while also demonstrating how technology may lead to positive social change, which is an important aspect to explore in our research.

## III. EXISTING SYSTEM

Individuals with disabilities (PWD) are constituents of every community. Chronic illnesses are one of the many factors contributing to the growing PWD population. The primary obstacles that people with disabilities encounter are obtaining appropriate education and underemployment. Particularly for full-time jobs, the percentage of employed persons with disabilities is lower than the percentage of employed individuals without impairments. [3]

In the existing system, the accessibility and usability of the page are hampered by the flaws in it. The fact that there are so many CVs on file and job openings listed on these platforms suggests that online hiring is succeeding in drawing in both employers and applicants. In terms of the content, aside from the general issues pertaining to every job, there are specific issues that are relevant to this particular audience. These include the accessibility of the company, the devices that the contractor permits, the use of filters based on the candidate's disability, and any special needs that the business may have. [8]

The current systems lack mechanisms or modules that particularly address the particular needs and preferences of job seekers with disabilities, even if they are frequently created with aesthetically pleasing and user-friendly interfaces. These systems are devoid of the necessary elements that enable disabled people to learn everything there is to know about the businesses they are considering. Furthermore, the current mechanisms are ineffective in helping recruiters and disabled job seekers communicate about their individual needs. While the current systems try to present a wide range of work prospects, they frequently fall short of prioritizing and filtering those that are truly in line with the talents and goals of job searchers with disabilities. As a result, a lot of disabled job searchers struggle to find appropriate occupations or have to accept positions that don't fit with their values and aspirations for their careers. This situation highlights the need for a novel approach that customizes the job search process to the particular requirements of job seekers with disabilities, giving them access to a platform that guarantees that the job opportunities displayed align with their goals, abilities, and the inclusive work environments they are looking for. [7]

## IV. PROPOSED SYSTEM

The kind of disability is important in terms of employment. The type of handicap and work capacity are strongly associated, so it is not surprising that work ability has an effect on career chances.[4] Getting hired can be hampered by a number of accessibility problems, such as inaccessible job applications, tests administered before and after the interview, and accommodations provided during the interview. Frequently, the hiring procedure for employees, which facilitates job seekers' application and hiring, may not be accessible to those with visual impairments.[6]

By offering an inclusive and user-friendly platform, the proposed online job portal for people with disabilities seeks to address the shortcomings of the current system. Including Voice-enabled text-to-speech functionality for users who are blind or visually impaired will be one of the main features. This is a summary of the suggested system:

Accessibility Features: UI elements, job listings, and navigation options can all be audibly described with the help of a comprehensive Voice-enabled text-to-speech feature. This will make it much easier for users who are blind or visually impaired to comprehend and use the platform.

Inclusive Content:
Reachable Job Listings: Make it mandatory for employers to submit job listings in an easily readable format that does not only rely on pictures.

Other Formats: Offer job descriptions in alternate formats, such as plain text in addition to graphics.

Education and Awareness:

Guides and Tutorials: To ensure that users can take full advantage of the platform's features, offer tutorials and guides on how to use accessibility features effectively.

There are several barriers which may arise when creating this project and some ways to mitigate are enhancing group training to incorporate web creation into the routine activities of service users, with a focus on self-advocacy and skill development, and fostering better collaboration between the project team and the involved sites through early communication to address logistical difficulties. Clearly defining the project's objective as a hub for dynamic knowledge exchange for all stakeholders involved in tackling the transition challenge is also important. [9]

## V. REQUIREMENT SPECIFICATIONS

Below are the technical requirements:

Previous: HTML, CSS, PHP
Background: MYSQL
Web Release: Python
IDE: Visual Studio Code

HARDWARE: Desktop with a minimum of 4GB of RAM (recommended 8GB)

## VI. RESULTS & DISCUSSIONS

### 1. VoiceOver

The purpose of the VoiceOver feature is to improve the platform's usability for individuals with visual impairments. Through the use of AI technology, it offers text-to-speech capabilities that let users interact and navigate the platform by hearing spoken feedback.

**Functionality:**

Generates speech from text content, such as user profiles, job listings, and platform instructions. Enables users to browse job listings, read job descriptions, and carry out other tasks with voice-guided navigation. Enables voice commands to be used by users to interact with the platform and perform tasks like searching, job applications, and profile updates.

**Benefits:**

Makes the platform more accessible, enabling people with visual impairments to use it. Enables users who are blind to look for jobs and scholarships on their own. Increases inclusivity by serving a larger number of users.

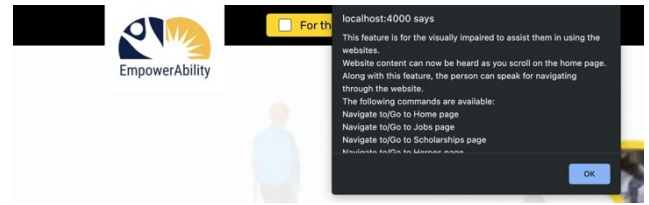

Fig.1 VoiceOver Feature

### 2. Home Page

The Home Page provides a user-friendly and educational landing page and acts as the user's point of entry.

**Functionality:**

Gives a summary of the goals and attributes of the platform. Provides easy access to important areas such as user profiles, success stories, job listings, and scholarship information.

**Benefits:**

Gives new users a warm and educational first impression. Makes it easier for users to navigate by providing direct links to important sections. Clearly visible calls to action that encourage user participation.

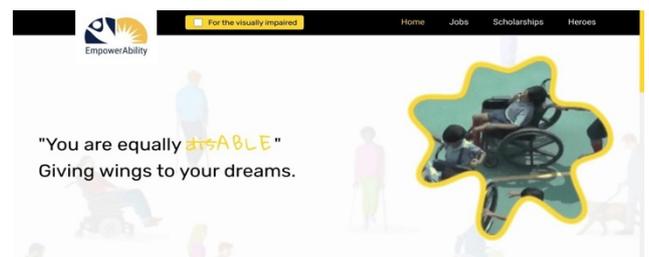

Fig.2 Home Page

### 3. Jobs Page

Users can browse and apply to jobs that are suited to their skills and preferences on the Jobs Page.

**Functionality:**

Displays job postings with job titles, company names, and locations in an orderly and user-friendly format. Enables users to filter job listings according to parameters like industry, job type, and location. Gives thorough job descriptions and options for applying to every posting. Allows users to access external application links or apply for jobs directly through the platform.

**Benefits:**

By consolidating a large number of job opportunities, it streamlines the job search process. Gives users the ability to filter and locate jobs based on their interests and skill set. Provides a simple application process that integrates seamlessly.

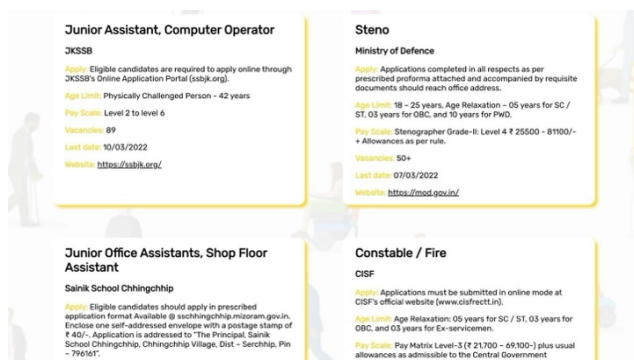

Fig.3 Jobs Page

### 4. Scholarships Page

Users can find resources to learn about funding opportunities for education by visiting the Scholarships Page.

**Functionality:**

Provides a list of available scholarships along with their names, requirements for eligibility, and deadlines for applications. Enables users to filter scholarships according to a number of factors, including study area and scholarship amount.Includes links and comprehensive details about each scholarship, including application requirements. Helps users apply for scholarships and monitors the status of their applications.

**Benefits:**

Makes it easier for people to access resources for financial aid, especially those who are pursuing education. Helps users find scholarships that fit their learning objectives. Simplifies the process of applying for scholarships.

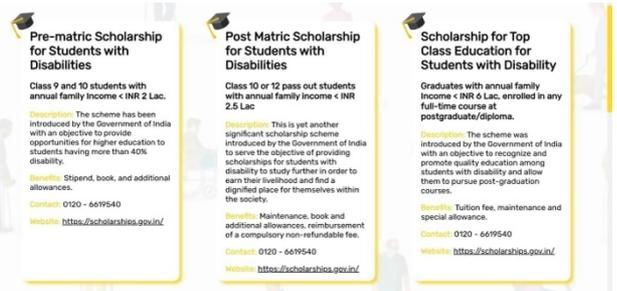

Fig.4 Scholarships Page

### 5. Heroes Page

The Hero Page features testimonials from people who have used the platform to overcome obstacles and succeed.

**Functionality:**

Features motivational testimonies and accomplishments of users who secured scholarships or jobs. Incorporates text, pictures, and video elements in addition to multimedia to tell these stories. Possibly includes endorsements and testimonies from users. Motivates users and cultivates a feeling of community.

**Benefits:**

Encourages and instills confidence in those applying for scholarships and jobs. Demonstrates the influence of the platform by presenting actual success stories. Fosters a sense of community and belonging among users.

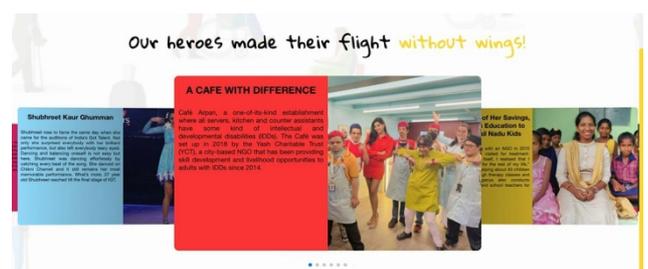

Fig.5 Heroes Page

### VII. CONCLUSION

According to a recent study on the obstacles experienced by a particular demographic of differently abled people in India, the majority of respondents reported facing significant physical, psychological, and ICT difficulties. While the majority of respondents (65%) indicated that there were significant physical hurdles in their lives, a sizable portion (52.5%) indicated that there were significant psychological and ICT barriers. These obstacles severely hinder their capacity to find employment opportunities. [10]

This project is a shining example of change in India's diverse landscape, utilizing cutting-edge technologies such as web scraping and AI-assisted speech-to-text features to foster inclusivity for people with special needs. With roughly 2 crores of people, or 2.1% of India's total population, the project tackles the urgent issue of why opportunities for people with special needs continue to be elusive.

This project is proof of the ability of technology to promote inclusivity. It goes beyond simply providing opportunities by boosting accessibility with cutting-edge features and inspiring through success stories. With continuous improvements including international expansion, AI-driven skill matching, and immersive learning opportunities, the platform has enormous potential to have a transformative impact. More than just a project, it is a mission that aims to rewrite the story for specially-abled people in India by combining technology and willpower in a more inclusive and equitable future. The project continues to transform lives and opportunities by going beyond the confines of disability.

## VIII. FUTURE ENHANCEMENTS

Peer support networks and mentorship programs should be enhanced with features that let users connect with peers who have gone through comparable struggles and seasoned mentors.

Cooperation with Educational Institutions: Establish alliances with colleges and technical schools to give students with special needs exclusive access to educational materials and assistance.

Certification and Skills Badges: To further improve users' employability, implement a certification program and skills badges that they can obtain through professional development and online courses.